\documentclass[twocolumn,nofootinbib,amsmath,prd,aps,superscriptaddress,tightenlines,preprintnumbers]{revtex4}
\pdfoutput=1

\usepackage{amsmath}
\usepackage{amssymb}
\usepackage{graphicx}
\usepackage{color}
\usepackage{tikz}
\RequirePackage[normalem]{ulem} %DIF PREAMBLE
\RequirePackage{color}\definecolor{RED}{rgb}{1,0,0}\definecolor{BLUE}{rgb}{0,0,1} %DIF PREAMBLE
\begin{document}

\newcount\hour \newcount\minute
\hour=\time \divide \hour by 60
\minute=\time
\count99=\hour \multiply \count99 by -60 \advance \minute by \count99
\newcommand{\mydate}{\ \today \ - \number\hour :00}
\newcommand{\andre}[1]{\textbf{\color{red} #1}}
\newcommand{\juan}[1]{\textbf{\color{blue} #1}}
\title{Time-dependent rate of multicomponent dark matter: Reproducing
	the DAMA/LIBRA phase-2 results}

%\title{Reproducing the DAMA/LIBRA phase-2 results with two dark matter components}
\def\coepp{ARC Centre of Excellence for Particle Physics at the Terascale, Department of
	Physics, University of Adelaide, Adelaide, South Australia 5005, Australia}

\author{Juan Herrero-Garcia}
\email{juan.herrero-garcia@coepp.org.au}

\author{Andre Scaffidi}
\email{andre.scaffidi@adelaide.edu.au}

\author{Martin White}
\email{martin.white@adelaide.edu.au}
\author{and Anthony G. Williams} 
\email{anthony.williams@adelaide.edu.au}

\affiliation{ARC Centre of Excellence for Particle Physics at the Terascale, Department of
	Physics, University of Adelaide, Adelaide, South Australia 5005, Australia}

\begin{abstract}
{The current paradigm for dark matter direct detection is to assume that the dark sector is solely composed
	of a single particle species. In this short paper, we make the observation that dark matter comprising both a
	light and a heavy component that modulate out of phase leads to interesting phenomenology in annual
	modulation experiments. For an illustrative example, we use the recently released DAMA/LIBRA phase-2
	results with a lower energy threshold. Immediately after, it was argued that a one-component spin-
	independent dark matter explanation of the observed annual modulation is strongly disfavored or excluded
	unless isospin-violating couplings are invoked. We show that a simple two-component extension can
	reproduce the observed spectrum without the need to invoke fine-tuned couplings. Using the publicly
	available DAMA/LIBRA data, we perform a fit of the DAMA/LIBRA energy spectrum of the annual
	modulation amplitude to a scenario with two dark matter components. We also take into account how
	gravitational focusing affects the phases of the light and a heavy components differently, which leads to
	nontrivial effects in the total time-dependent rate. Our results show that there exists a unique solution in
	agreement with the data in the simplest case of isospin-conserving couplings with equal cross sections. The
	distinctive features found in this work are crucial for a dark matter interpretation of any observed annual
	modulation.}

\end{abstract}

\maketitle

%%%%%%%%%%%%%%%%%%%%%%%
\section{Introduction} \label{sec:intro}
%%%%%%%%%%%%%%%%%%%%%%%%

Dark matter (DM) is one of natures greatest enigmas. Until now evidence for its existence stems from its gravitational interactions. {However, it is by no means guaranteed that just a single state or particle (1DM) constitutes the whole dark sector, which may have a multi-component nature similarly to the visible sector.}  Multi-component DM in direct detection has only been studied in a few works which have then only focused on time-averaged (not modulated) event rates~\cite{Profumo:2009tb,Batell:2009vb,Adulpravitchai:2011ei,Dienes:2012cf,Chialva:2012rq,Bhattacharya:2016ysw,Bhattacharya:2017fid,Herrero-Garcia:2017vrl}. A generic prediction of two-component DM (2DM) is the presence of kinks in the differential event rates in mono-target experiments due to the different DM components~\cite{Herrero-Garcia:2017vrl}. {In the following study, we observe another interesting prediction of 2DM in
annual modulation experiments. Namely, that a
light and heavy component will modulate out of
phase, producing non-trivial modulation ampli-
tudes that significantly affect the interpretation of the results.} {To study this effect,} we adopt a purely phenomenological approach to try to reproduce the modulated signal {observed by the DAMA/LIBRA collaboration~\cite{Bernabei:2010mq,Bernabei:2013xsa} (denoted by DAMA in the following).} {We do so} without going into further details regarding the model building, which would affect the interactions anda the abundances of the DM components. This is in fact studied for averaged rates in Ref.~\cite{Herrero-Garcia:2018qnz}.  In this work we also include gravitational focusing (GF)~\cite{Alenazi:2006wu,Lee:2013wza,Bozorgnia:2014dqa}, which has a non-trivial effect on the 2DM time dependence of the direct detection rates, and therefore on the modulation amplitude and phase. Although in our work we use the DAMA data, the physics discussed and the results obtained in the following apply to generic time-dependent signals of multi-component scenarios.

Very recently, the DAMA collaboration released the long-waited phase-2 results with a lower energy threshold, with two new energy bins below 2 keVee~\cite{DAMAphase2,Bernabei:2018yyw}. As first pointed out in version 3 of Ref.~\cite{Kahlhoefer:2018knc}, and studied in Ref.~\cite{Baum:2018ekm} (see also Ref.~\cite{Kang:2018qvz}), the consistency of the DM interpretation of DAMA's signal is now under question both for the light and the heavy DM mass solutions mentioned above for vanilla isospin-conserving spin-independent (SI) interactions, even before considering its compatibility with other null-result experiments.  This is because below 2 keVee, the direct detection rates for the two standard DM solutions behave very differently with decreasing recoil energy: the light DM gives rise to scatterings off iodine, increasing its rate significantly, while for the heavy DM  (that scatters predominately off iodine) the modulation amplitude decreases, eventually giving rise to a phase flip. This was already pointed out in Ref.~\cite{Kelso:2013gda}. We confirm in this work that for SI interactions the energy spectrum for the one-component DM scenario (1DM) is disfavoured. Indeed we find that the heavy solution is excluded at an even higher significance level when GF effects are accounted for.

The current significance for the modulation in the complete DAMA data set is $12.9\,\sigma$. The DAMA collaboration, as well as several independent studies, have not found that modulated backgrounds like those of neutrons, muons, solar neutrinos or radon can be responsible for the signal~\cite{Chang:2011eb,FernandezMartinez:2012wd,Klinger:2016tlo,Belli:2017hqs,Bernabei:2018drd,McKinsey:2018xdb}. Until the very latest phase-2 results, under reasonable particle physics and astrophysical assumptions, the signal was consistent in both amplitude and phase with that expected from Weakly Interacting Massive Particles (WIMPs). Assuming the Standard Halo Model (SHM) and elastic scattering, the best-fit masses (and cross sections) are well-known: a light DM with mass $\sim 10$ GeV scattering mainly on sodium (light mass solution), or a heavy DM with mass $\sim 70$ GeV scattering mainly on iodine (heavy mass solution)~\cite{Bottino:2003cz,Bottino:2003iu,Gondolo:2005hh,Fairbairn:2008gz,Kopp:2009qt, Schwetz:2011xm, DelNobile:2015lxa,Herrero-Garcia:2015kga}. We show that the modulation observed by DAMA at low energies can be reproduced in a natural way by a combination of two DM particles (2DM), without the need to invoke fine-tuned isospin-violating couplings\footnote{We note that {there have been studies that show that other effective operators can relieve the DAMA self-tension~\cite{Kang:2018qvz}. Also the same authors claim that compatibility with
other null results can be achieved in a proton-philic spin-dependent inelastic scenario when the
DM velocity distribution departs from Maxwellian~\cite{Kang:2018zld}.}}.

The main issue with a DM interpretation of the DAMA modulation is that there is currently no accepted explanation that reconciles DAMA's signal with the absence of a positive signal in all other experiments~\cite{Akerib:2016vxi,Aprile:2017iyp,Cui:2017nnn,Savage:2010tg,Schwetz:2011xm}, even independently of the DM velocity distribution~\cite{McCabe:2011sr,Frandsen:2011ts,Frandsen:2011gi,HerreroGarcia:2011aa,HerreroGarcia:2012fu,Bozorgnia:2013hsa,Gelmini:2016pei}. This has motivated a large experimental effort to try to reproduce the DAMA experiment with NaI crystals in order to independently either confirm or reject its results ~\cite{Amare:2014jta,Fushimi:2015sew,Shields:2015wka,Angloher:2016ooq,Thompson:2017yvq}. The SABRE experiment~\cite{Shields:2015wka} plans to have a northern and southern hemisphere pair of NaI detectors to search for a seasonal correlation or anti-correlation of any DAMA-like modulation signal~\cite{Froborg:2016ova}. Interestingly, the COSINUS experiment~\cite{Angloher:2016ooq} aims to also measure the constant rate by developing a cryogenic detector. A null result in the latter experiment may rule out a DM explanation of DAMA model-independently~\cite{Kahlhoefer:2018knc}.

This letter is structured as follows. In Sec.~\ref{sec:annualmod} we introduce the relevant notation to describe the DM time-dependent signal in direct detection experiments. In Sec.~\ref{sec:results} we fit the binned amplitude of the DAMA modulation. We do this by generating pseudo-mock data from a modulating 2DM signal, including GF, and extracting the modulation amplitude by fitting a sinusoid to the resulting time-dependent signal. This allows us to draw conclusions regarding the effect of GF and the non-sinusoidal component. In this section we also conduct the analysis for a single light/heavy DM to compare with results from other studies. We give our conclusions and final remarks in Sec.~\ref{sec:conc}.

\section{The dark matter direct detection signal}\label{sec:annualmod}

In this section, we present the relevant expressions for the direct detection of 2DM, with individual DM masses $m_{1,2}$ (we take $m_1<m_2$), SI cross-sections with protons $\sigma_{1,2}^p$, and local energy densities $\rho_{1,2}$. We use $\rho_1+\rho_2=\rho_{\rm loc}$, where $\rho_{\rm loc}$ is the local DM mass density. We take the astrophysical values of Ref.~\cite{Herrero-Garcia:2017vrl}, and use  the notation $r_{\rho} \equiv \rho_2/\rho_1$ and $r_{\sigma} \equiv \sigma_2^p/\sigma_1^p$. 

For 2DM with SI interactions, we can parameterise the differential event rate produced on a target nucleus with mass number A as~\cite{Herrero-Garcia:2017vrl}:
\begin{align} 
\label{eq:rate_tot}
\mathcal{R}_{A}(E_{R},t) &=  \frac{x_A\,\rho_{\rm loc}\,\sigma_1^p}{2 \,(1+r_\rho)\,\mu_{p1}^2} \,A^2 F_{A}^2(E_{R}) \\
&\times\left[\frac{\eta(v_{m,A}^{(1)},t)}{m_1}+\frac{r_{\rho}r_{\sigma}\mu_{p1}^2}{\mu_{p2}^2}\frac{\eta(v_{m,A}^{(2)},t)}{m_2}\right] \nonumber\,,
\end{align}
where here we define the halo integral as
\begin{equation}\label{eq:eta} 
\eta (v_{m,A}^{(\alpha)},t) =
\int_{v > v_{m,A}^{(\alpha)}} \negthickspace \negthickspace d^3 v \,\frac{f^{(\alpha)}_{\rm det}(\vec{v},t)}{v} \,.
\end{equation}
From kinematics, the DM particle $\alpha$ ($\alpha=1,2$) is required to have a velocity larger than $v_{m,A}^{(\alpha)}=\sqrt{m_A E_{R}/(2 \mu_{\alpha A}^2)}$ to produce a recoil of energy $E_R$. $m_A$ is the mass of the nucleus, $x_A$ its mass faction in the detector, $\mu_{\alpha A}$ is the reduced DM-nucleus mass and $F_A(E_{R})$ is its nuclear form factor, for which we use the Helm parametrisation~\cite{LEWIN199687,PhysRev.104.1466}. In Eq.~\eqref{eq:eta} $f^{(\alpha)}_{\rm det}(\vec v, t)$ describes the distribution of DM particle velocities in the detector rest frame.  In the following we use the SHM, i.e., a Maxwellian velocity distribution, with equal velocity dispersions for both DM components (see Ref.~\cite{Herrero-Garcia:2018qnz} for a study regarding this assumption). As before, the total rate is given by the sum of the contributions on each target nucleus, i.e., $\mathcal{R}(E_{R},t)=\sum_A \mathcal{R}_{A}(E_{R},t)$. The rate for 1DM is obtained from Eq.~\eqref{eq:rate_tot} in the limit $r_\rho=0$. 

In this work we deal with the time-dependent signal caused by the motion of the Earth around the Sun~\cite{Drukier:1986tm, Freese:1987wu}. The total differential event rate can be decomposed as (see for instance Refs.~\cite{HerreroGarcia:2011aa,Freese:2012xd,HerreroGarcia:2012fu,Herrero-Garcia:2015kga})
\begin{align}
\label{eq:rate_tot_t}
\mathcal{R}^{(\alpha)}_{A} (E_R, t) \equiv \bar R^{(\alpha)}_{A}  (E_{R})  + \mathcal{M}^{(\alpha)}_{A} (E_{R},t)\,,
\end{align}
where $\bar R^{(\alpha)}_{A}$ is the time-averaged rate and $\mathcal{M}^{(\alpha)}_{A} (E_{R},t)$ the time-dependent signal. In the following we assume that all the time-dependence stems from the velocity of the Earth $v_e(t)$, which is a reasonable assumption for the time scales of direct detection experiments. For isotropic and sufficiently smooth DM haloes, to leading order on $v_e(t)$,\footnote{A nonsinusoidal modulation is expected when higher-order
	harmonics, which are, however, suppressed by extra powers of
	$v_e$ , are considered~\cite{Chang:2011eb,Lee:2013wza,Bozorgnia:2014dqa}.} the time-dependent signal consists of the annual modulation, 
\begin{align}
\label{eq:AM_t}
\mathcal{M}^{(\alpha)}_{A} (E_{R},t) = \mathcal{M}^{(\alpha)}_{A} (E_{R}) \cos[2\pi (t-t_0(v_{m,A}^{(\alpha)})]\,,
\end{align}
where $t_0(v_{m,A}^{(\alpha)})$ is the phase of the modulation and 
\begin{equation}\label{eq:Aeta} 
\mathcal{M}^{(\alpha)}_{A} (E_{R})= \frac{1}{2}\left[ \mathcal{M}^{(\alpha)}_{A} (E_{R},t_{\rm J})-  \mathcal{M}^{(\alpha)}_{A} (E_{R},t_{\rm J}+0.5) \right] \,,
\end{equation}
is the amplitude. $t_{\rm J}  \;(t_{\rm J}+0.5)$  measured in years corresponds to June (December) 1st, which for minimum velocities $v_{m,A}^{(\alpha)}$ above $\sim 200$\,kms$^{-1}$ is the time of the year when the velocity of the WIMP flow in the Earth's frame is maximum (minimum). For $v_{m,A}^{(\alpha)} <200$\,kms$^{-1}$, the situation is the opposite: $t_{\rm J}\,(t_{\rm J}+0.5)$ $t_{\rm J}\;(t_{\rm J}+0.5)$ corresponds to minimum (maximum) WIMP flow, such that $\mathcal{M}_{A}^{(\alpha)}(E_R)$ becomes negative. In other words, for small enough $v_{m,A}^{(\alpha)}$, the phase of the modulation flips by 6 months. As we show below, this is precisely what happens for heavy DM components scattering off iodine in DAMA at the lowest energies. The total rates are given by the sum over nuclei and DM components, $\bar R(E_{R})=\sum_{\alpha, A} \bar R^{(\alpha)}_{A}(E_{R})$ and $\mathcal{M}(E_{R},t)=\sum_{\alpha, A} \mathcal{M}^{(\alpha)}_{A}(E_{R},t)$.

\subsection{Gravitational focusing and the non-sinusoidal signal}  \label{sec:grav_foc} 

GF of DM particles by the Sun affects the phase of the modulation $t_0$, which, for an observed $E_R$, is sensitive to the DM mass via the dependence on $v_{m,A}^{(\alpha)}$~\cite{Alenazi:2006wu,Lee:2013wza,Bozorgnia:2014dqa}. The effect is significant for $v_{m,A}^{(\alpha)} \lesssim 200\, {\rm km\,s}^{-1}$, in which case the phase is changed from the vanilla case of December 1st towards a later value. In the case of DAMA the effect on the phase is important for heavy DM masses, for which it can change by tens of days. For light DM particles, the effect is negligible and the phase remains at June 1st. 
	In 2DM the most interesting feature is that the sum of the time-dependent signals of the individual components, with their phases $t^{(1,2)}_0$ being different due to GF ($t^{(2)}_0 > t^{(1)}_0$ as $m_2>m_1$), leads to a non-sinusoidal time-dependent signal only suppressed by the (small) phase difference of the 2DM components $\Delta t_0\equiv t^{(2)}_0-t^{(1)}_0 > 0$. 

	We can see this by expanding the combined signal in $\Delta t_0$ (see also Ref.~\cite{Freese:2012xd}). For a given nuclei and recoil energy, $\Delta t_0$ depends on the mass splitting of both DM components. We find that the combined $\mathcal{M} (E_{R},t)$ is given to leading order in $\Delta  t_0$ by
	\begin{align}
	\label{eq:rate_tot_GF}
	\mathcal{M}(t) &=\mathcal{M}^{(1)}\cos[2\pi(t-t^{(1)}_0)]  + \mathcal{M}^{(2)}\cos[2\pi(t-t^{(1)}_0-\Delta t_0)] \nonumber\\
	&=(\mathcal{M}^{(1)}+  \mathcal{M}^{(2)})\,\cos[2\pi(t-t^{(1)}_0)] \nonumber\\
	&+ \mathcal{M}^{(2)}\,\left(\frac{\Delta t_0}{t-t^{(1)}_0}\right) \,\sin[2\pi(t-t^{(1)}_0)] \,,
	\end{align}
	where the expansion is valid for $t\gg t^{(2)}_0> t^{(1)}_0$. For $t\sim t^{(1,2)}_0$ higher order terms become important. As can be observed, the sine term is proportional to $\Delta t_0$. Therefore, for non-negligible $\Delta t_0 $, the total time-dependent signal is non-sinusoidal. For the DAMA observation, how good this approximation is depends on the DM masses, which we study in the following by fitting the DAMA-phase 2 time-dependent signal to the numerically computed one using Eq.~\eqref{eq:rate_tot} with GF effects implemented.

\section{The DAMA energy spectrum of the modulation amplitude} \label{sec:results}
We perform our analysis on the whole energy spectrum of the DAMA annual modulation amplitude, which combines results from DAMA/NaI and DAMA/LIBRA phases 1 and 2. The total exposure is 2.46 kg$\cdot$y. The events are measured in electron equivalent energy (keVee), which is related to the true recoil energy $E_R$ through the target-dependent quenching factors $E_\text{ee} = Q_A\,E_R$ (we use $Q_{\rm Na}= 0.3$, $Q_{\rm I} = 0.09$). We employ the differential response function of Ref.~\cite{Bernabei:2008yh}, treating the parameters $\alpha_{\rm LE} =(0.448 \pm 0.035)\,\sqrt{\text{keVee}}$ and $\beta_{\rm LE}=(9.1 \pm 5.1)\times 10^{-3}$ as nuisance parameters.
The results made public by the DAMA collaboration are presented in slide 22 of Ref.~\cite{DAMAphase2}. We use the data of Tab.~I of Ref.~\cite{Baum:2018ekm}, which gives the observed binned annual modulation amplitude $M_i$ in $N=10$ bins in the energy range [1, 20] keVee in order to increase the signal to noise ratio in our fit.

We undertake our analysis as follows:
	 \begin{enumerate} 
		\item 	We first generate pseudo-mock data from a modulating DM signal with gravitational focusing corrections implemented and $2.46\,\text{ton\,y}$ of exposure in the time intervals used by DAMA. The term `pseudo-mock' here refers to the fact that we use the `Asimov data', which in the large statistics limit corresponds to the expected Poisson mean in each time bin. This is done to ensure smooth likelihood functions in our statistical analyses. We only generate signal events, since the DAMA collaboration releases no information about the observed constant backgrounds.
		\item As DAMA does, we then fit the function $R_A(t)=S_0+S_m\cos[2\pi(t-t_0)/T]$ to the pseudo-mock data set in each energy bin.  Since the DAMA experiment does not veto backgrounds, we fit a sinusoidal component plus a constant offset to extract the best fit modulation amplitude $S_m$. We take a constant phase of $t_0=152.5$ days and a period $T=1\,$yr.\footnote{The DAMA collaboration also presents results assuming variable phases and additional non-sinusoidal components. Energy spectra for these fits however are presented in larger integrated energy bins of $1$ keVee width. Since any interesting behaviour in the spectrum occurs in the lowest $[1-1.5]$ keVee energy bin we do not use these other data sets.} 
		\item    We take the estimated amplitude for each energy bin and compare these to the DAMA data using a maximum likelihood analysis in order to extract the goodness of fit and best-fit values of the 1DM and 2DM model parameters. 
	 \end{enumerate} 
The reason for this roundabout approach is that the DAMA collaboration does not release time-dependent data in 0.5 keVee bins,\footnote{The only time-dependent information released by the DAMA collaboration are residuals in large integrated energy bins [1-3], [2-6] and [1-6] keVee. However, in the latter the energy information of the (most interesting) lowest energy bins is washed out.} and so we proceed as closely as possible to the method used by the collaboration to extract the modulation amplitude. This would not be necessary if we did not consider GF, but GF  non-trivially changes the phase of the modulation away from June 1st as well as the shape of the time-dependent signal.

We use the open source software {\tt Minuit}~\cite{James:1994vla}  to compute the best-fit points, which we give in Tabs.~\ref{tab:fit1DAMASI} and \ref{tab:fit2DAMA}.  For completeness, we also compute the corresponding p-values for the fits, as well as the corresponding number of equivalent two-sided Gaussian standard deviations $\mathcal{Z}$. We deem a `good fit' to be one that gives a p-value $p>0.05$ which corresponds to $\mathcal{Z}<1.96$.  For 2DM, we also employ the {\tt MultiNest} implementation of the nested sampling algorithm~\cite{Feroz:2008xx,Feroz:2007kg,Feroz:2013hea}, with $5000$ live points and a tolerance of 0.5. We then determine the distribution of the profile likelihood ratio (PLR) $\mathcal{L}/\mathcal{L}_\text{max}$ throughout the parameter space from the obtained samples.
\begin{table}[t!]
	\small
	\centering
	\begin{tabular}{ccccccccc}
		1DM& $m$ (GeV) &$\sigma_1^p$ ($\times10^{-40}\,{\rm cm}^2$)  & $\chi_{\rm min}^2/{\rm dof}$ & p-value & $\mathcal{Z}$ \\ \hline\hline
		Light &  8.42 & 1.25& 48.4/8& $8.15\times 10^{-8}$& 5.40 \\\hline
		Heavy &  70.1 & 0.10& 56.8/8& $1.94\times 10^{-9}$ & 6.00 \\\hline \hline
		\end{tabular}
	\caption{ Results of 1DM fit to the DAMA energy spectrum after accounting for GF corrections. The different fits are for both the light and the heavy DM mass solutions. The dashes refer to the parameters that are fixed to 1.}
	\label{tab:fit1DAMASI}
\end{table}
\begin{table}[h!]
	\centering
	\begin{tabular}{cccccccccccc}
		2DM& $m_1$ &$m_2$ &$\sigma_1^p$ &$r_\rho$ & $\chi_{\rm min}^2/{\rm dof}$ & p-value & $\mathcal{Z}$ \\ \hline\hline
		$r_\sigma=1$&  22.0 &80.3& 0.14& 3.35&11.8/6&$0.07$&1.84 \\\hline
	\end{tabular}
	\caption{Same as Tab.~\ref{tab:fit1DAMASI} for 2DM fit to the DAMA energy spectrum. }
	\label{tab:fit2DAMA}
\end{table}

\subsection{One-component dark matter fit}  \label{sec:fit1DM}

In Tab.~\ref{tab:fit1DAMASI} we show results of 1DM fits for the light and the heavy DM mass solutions, which correspond to scatterings mainly in Na and I, respectively. We also check the exclusion significances for the 1DM case neglecting gravitational focusing effects to check consistency with previous studies. Since GF does not effect the sinusoidality of the DAMA signal for DM masses below $\sim30$ GeV,  we find that the light solution is excluded at $5.4\sigma$, which is roughly the same significance as observed in previous studies. Interestingly, however, we see that including GF corrections non-trivially increases the exclusion significance of the heavy solution from $\sim3\sigma$ to $6\sigma$.
\begin{figure}[tb]  
	\centering
	\includegraphics[scale=0.28]{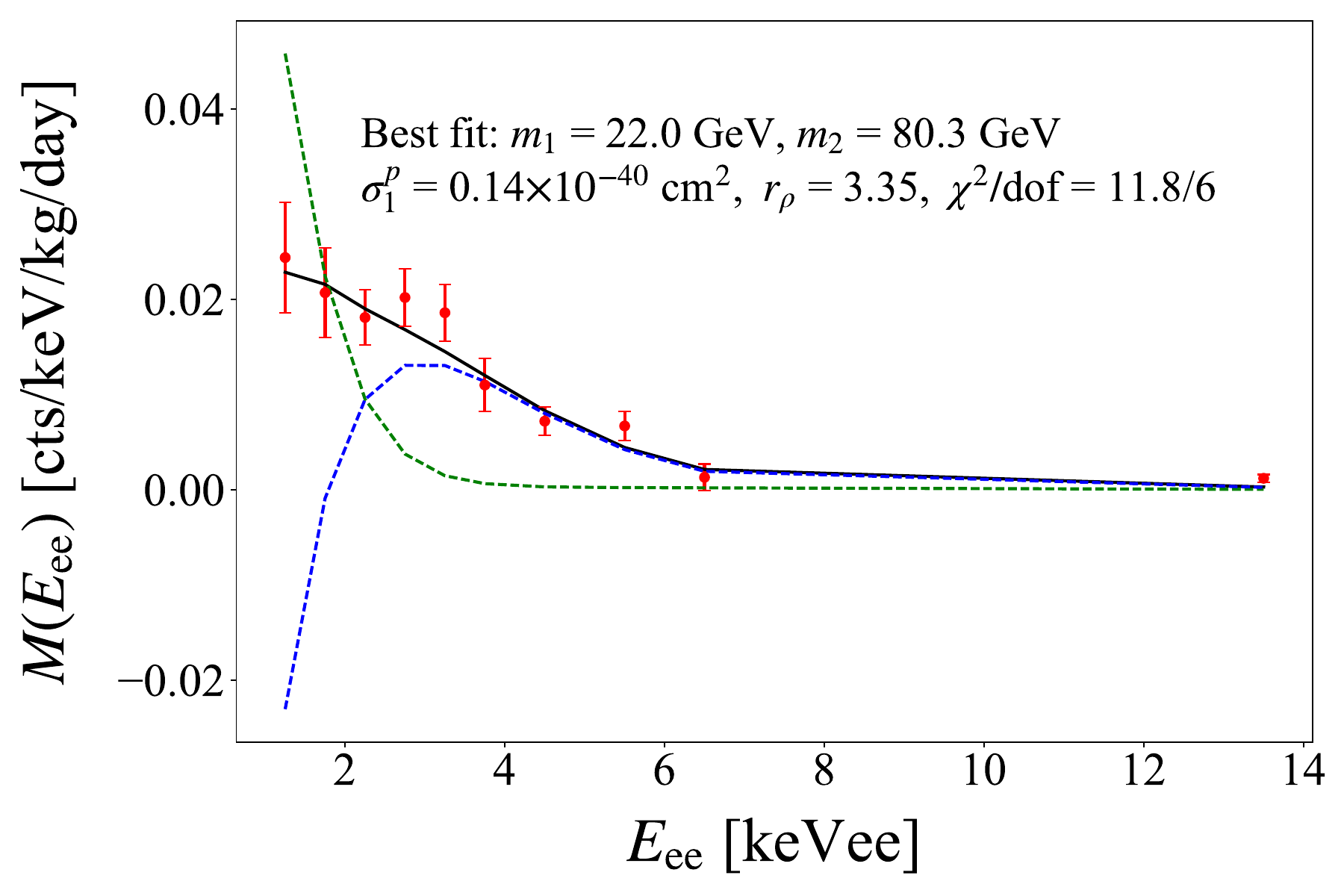}\\
	\vspace{0.3 mm}
	\includegraphics[width=0.485\textwidth,height=0.3\textwidth,keepaspectratio]{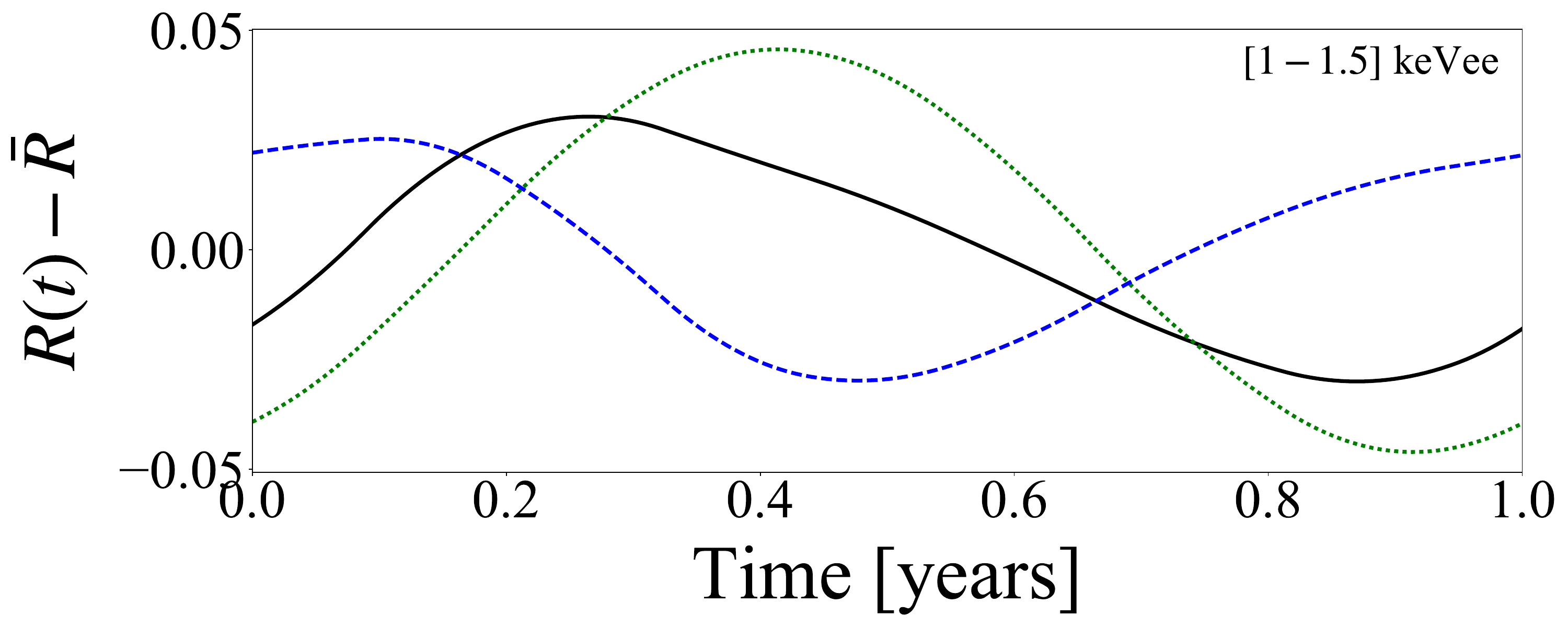}
	\caption{
		Two-component fit to the DAMA/LIBRA phase-2 data. We also show the contribution from DM1 (DM2) in \emph{dotted green} (\emph{dashed blue}) and the combined (\emph{solid black}). \emph{Upper panel:} Best-fit spectra of the DAMA energy spectrum for the 2DM model. The experimental best-fit modulation amplitudes to the combined DAMA phases 1 and 2 rebinned~\cite{Baum:2018ekm} are shown with \emph{red points}. Below the phase flip, which occurs at $\sim2$ keVee, the contributions of the two components partially neutralise each other in the combined modulation. 
		\emph{Lower panel:}  Time-dependent residuals in the lowest energy bin, [1-1.5] keVee. Notice the anti-modulation of DM2, and the non-sinusoidality of DM2 and the combined signal.}\label{fit:2DM}
\end{figure} 

\begin{figure*}[ht]
	\def  \path{4D_Ann_Mod_cos_GF_KF_out}
	\def \tail{_like2D}
	\def \sc{0.6}
	\centering
	\begin{tikzpicture}
	\node (letter) {\includegraphics[scale=\sc]{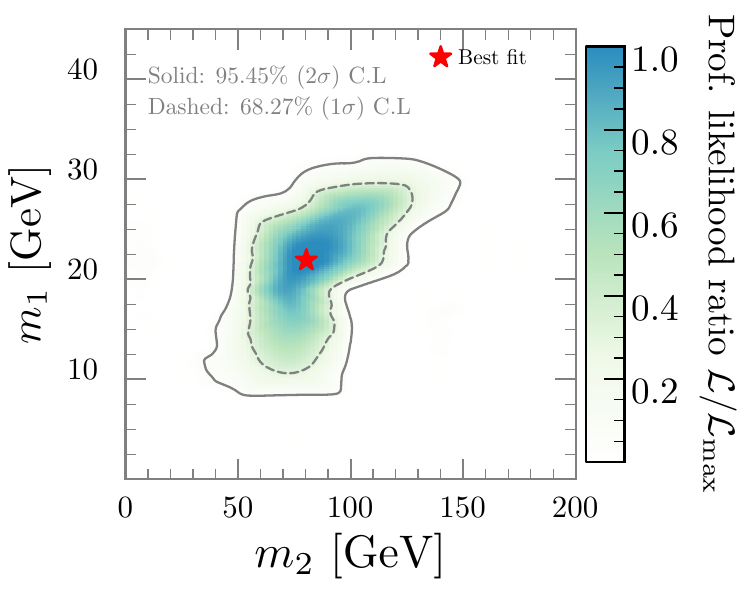}};
	\end{tikzpicture}~
	\begin{tikzpicture}
	\node (letter) {\includegraphics[scale=\sc]{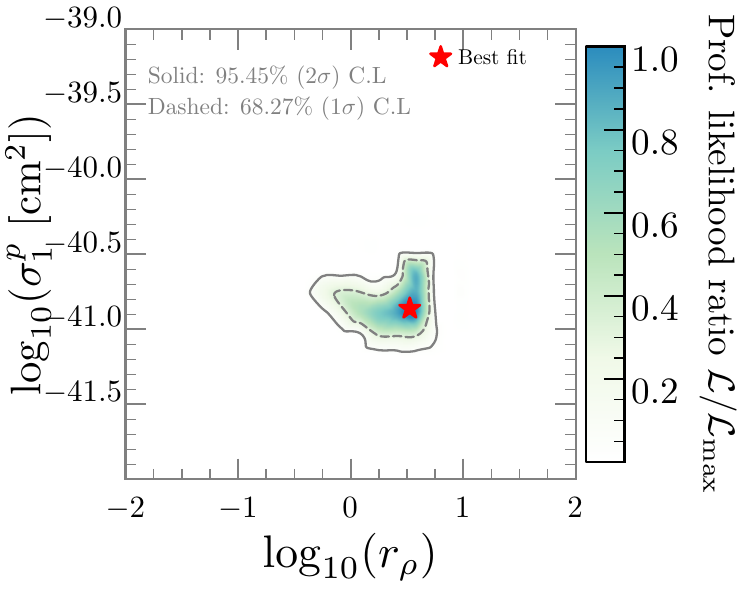}};
	\end{tikzpicture}\\
	\begin{tikzpicture}
	\node (letter) {\includegraphics[scale=\sc]{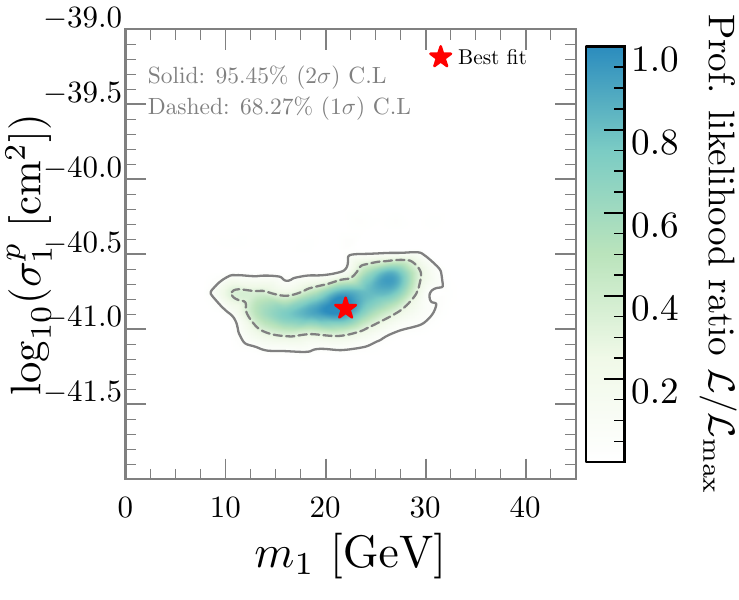}};
	\end{tikzpicture}~
	\begin{tikzpicture}
	\node (letter) {\includegraphics[scale=\sc]{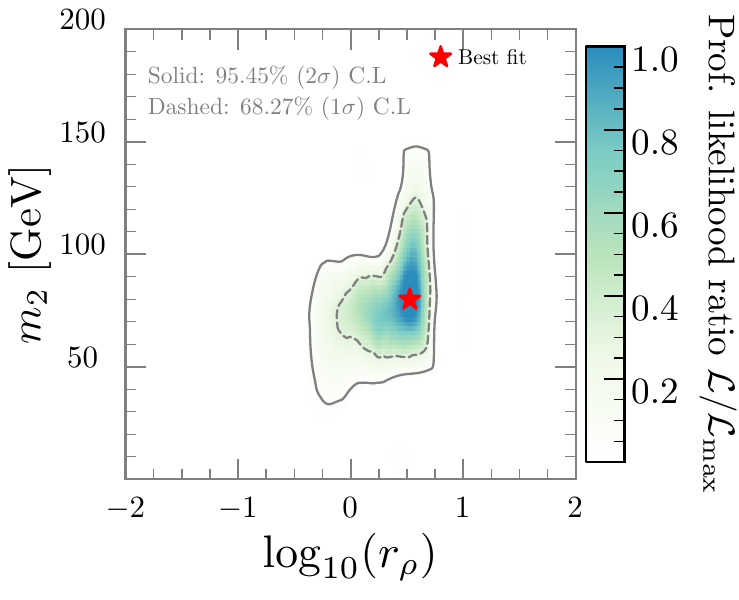}};
		\end{tikzpicture}%
	\caption{Profile likelihood ratio (PLR) density overlaid with 1 and 2$\sigma$ C.L. contours for the 2DM fit to the DAMA energy spectrum of the modulation amplitude as calculated in Sec.~\ref{sec:fit2DM}. We indicate the best fit points with a red star (see Tab.~\ref{tab:fit2DAMA}.)} \label{regions:4D}
\end{figure*}

\subsection{Two-component dark matter fit}  \label{sec:fit2DM} 
We show in Tab.~\ref{tab:fit2DAMA} the result of the fit for two DM particles (2DM). We fix $r_\sigma=1$, leaving $r_\rho$ free.\footnote{The energy density and the cross section enter identically in the rate. Therefore, apart from the overall normalisation, one can fix $r_\sigma$ and consider $r_\rho$ as a free parameter.} One can observe that the 2DM model provides a good fit with an exclusion significance of only $1.84\sigma$.
In Fig.~\ref{fit:2DM} (upper panel) we show the binned modulation amplitude (solid black) for the best-fit points of the 2DM solution. This corresponds to scattering of both DM components \emph{dominantly} off iodine. The fact that the lighter component scatters dominantly off iodine with a negligible contribution from sodium is due to two factors: first, the smaller quenching factor in iodine compensates its larger mass, translating into smaller $v^{(1)}_{\rm m,I} (E_R)$ and thus into larger $\eta(v^{(1)}_{\rm m,I})$; second is the $A^2$ enhancement factor for iodine due to coherent SI scatterings.  Below $2$ keVee DM2 becomes negative (i.e., a phase flip) and DM1 dominates the total rate. Therefore there is a partial cancellation in the combined modulation between the individual DM contributions. This is illustrated in Fig.~\ref{fit:2DM} (lower panel), where we plot the time-dependent residual for 2DM in the lowest recoil energy bin [1-1.5] keVee.\footnote{For completeness, we have also conducted a similar analysis without including GF. In this case, we obtain \textit{two} solutions. One which corresponds to a similar one as with GF, and another one at $m_1\sim8$ GeV and $m_2\sim170$ with $r_\rho\sim0.07$. That is, the second solution prefers a suppressed heavy component, i.e., a 1DM scenario.}

In Fig.~\ref{regions:4D} we present the frequentist 1 and 2$\sigma$ C.L. contours using {\tt pippi}~\cite{Scott:2012qh}. We show the planes $m_1$--$m_2$ (\emph{top left}), $\log_{10}(r_\rho)$--$\log_{10}(\sigma_1^p)$ (\emph{top right}), $\log_{10}(r_\rho)$--$m_2$ (\emph{bottom left}) and $\log_{10}(r_\rho)$-$m_2$ (\emph{bottom right}). The regions are very well-defined around the best-fit values. The $2\sigma$ range of $m_1$ is quite narrow, [8, 30] GeV, while that of $m_2$ is larger, [30, 150] GeV, i.e., there is more freedom in the heavy component since for heavier DM the dependence on the mass via $\eta(v_{m,A}^{(\alpha)})$ is milder. For large mass values, $m_1$ and $m_2$ show a mild positive correlation, such that both individual terms in the square bracket of Eq.~\eqref{eq:rate_tot} become similar in size. Moreover, one can observe how the largest $m_2$ region corresponds to large $r_\rho$ (\emph{bottom right} panel), which is easily understood looking at the second term in the square bracket of Eq.~\eqref{eq:rate_tot}. Also, the region of large $\sigma_1^p$ corresponds to large $r_\rho$,  see \emph{top left} panel, as expected from the overall normalisation of the rate.

\section{Conclusions} \label{sec:conc} 
{In this letter, we have studied how a simple 2DM can lead to interesting signals in annual modulation experiments. Furthermore, we showed that implementing corrections due to GF leads to non-sinusoidal and non-trivial phase effects in the evaluation of the 2DM time-dependent rate. As an illustrative example, we} studied whether a 2DM model comprising both a light and heavy component can revive the vanilla DM interpretation of the low threshold DAMA data. We performed a fit of 2DM to the publicly available DAMA data using the energy spectrum of the amplitude of the modulation. For the first time, we also fully incorporate gravitational focusing effects into such an analysis. Firstly, we fit 1DM to the energy spectrum data and find that the heavy solution is even more excluded than shown in previous studies after GF effects are considered. On the contrary, we find that 2DM provides very good agreement to the energy spectrum data. 

We show our solution in Fig.~\ref{fit:2DM}, where scatterings are predominantly off iodine, with a crossing between the spectrum of the two individual DM components at roughly 2 keVee. The crossing occurs due to a phase flip in the heavier component at low recoil energy. The results of the 2DM fits are summarised in Tab.~\ref{tab:fit2DAMA} and Figs.~\ref{fit:2DM} and \ref{regions:4D}, which involve reasonable values of the relative energy densities and the cross section. The key feature found in the analysis is that, at the lowest energies, the modulation of the heavy DM particle becomes negative, so that in the combined modulation the individual DM contributions partially cancel each other. 

Therefore two-component DM looks like a natural solution to the first part of the DAMA puzzle: the compatibility of the spectrum with that expected from DM under standard astrophysical and particle physics assumptions. The second part of the DAMA puzzle, that is, the compatibility of DAMA data with other null results is not solved. 

We would like to conclude by saying that it would be very helpful if the DAMA collaboration made public the temporal data in smaller energy bins, which would allow the use of all information. In particular it would show whether or not there is a non-sinusoidal behaviour in the current data in the [1,1.5] keVee bin, as present in the case of 2DM (see lower panel of Fig.~\ref{fit:2DM}). 

In any case, even if the DAMA signal turns out not to be related to DM, the distinctive features of the time-dependent signal of multi-component DM found in this work, like the non-sinusoidal behaviour and the possibility of having a partial cancellation at low energies, are generic predictions that should be searched for in case an annual modulation signal is observed in next generation experiments.

\vspace{0.7cm}
{\bf Acknowledgements:}
 JHG would like to thank Thomas Schwetz for useful discussions. This work is supported by the Australian Research Council through the Centre of Excellence for Particle Physics at the Terascale CE110001004. MW is supported by the Australian Research Council Future Fellowship FT140100244. MW wishes to thank Lucien Boland and Sean Crosby for their administration of, and ongoing assistance with, the MPI-enabled computing cluster on which this work was performed.

\vspace{1cm}

%%%%%%%%%%%%%%%%%%%%%%%%%%%%%%%%%%%%%%%%%%%%%%%%%%%%%%%%%%%%%%%%%%%%%%%%%%%%5
\bibliographystyle{my-h-physrev}

\bibliography{2DM_DAMA}

\end{document}